\documentclass[onecolumn,preprint,preprintnumbers,amsmath,amssymb]{revtex4}

\usepackage{graphicx}
\usepackage{dcolumn}
\usepackage{bm}

\begin{document}
\title{Vortex interactions in presence of a soft magnetic film}

\author{Lars Egil Helseth}
\affiliation{Max Planck Institute of Colloids and Interfaces, D-14424, Potsdam, Germany }%

\begin{abstract}
We study theoretically the behavior of vortices in a thin film superconductor placed close to 
a soft magnetic film. It is shown that the field from the vortex induces a magnetization 
distribution in the soft magnetic film, thus modifying the fields and vortex interactions. 
We suggest that the interaction between two otherwise identical vortices is attractive 
at short distances, but repulsive at larger distances. This is in contrast to the case without 
the soft magnetic film, where the force is always repulsive.

\end{abstract}

\pacs{Valid PACS appear here}
\maketitle
The structure of vortices in thin superconducting films was first investigated in detail by 
Pearl\cite{Pearl}. He found that such vortices interact mainly via their 
stray field, which extends far into the nonsuperconducting medium. Later, this
theory has been extended to thin film systems with and without 
anisotropy\cite{Clem,Carneiro,Clem1}. More recently, the interaction between 
vortices and magnetic nanostructures has gained significant interest, due to the possible 
enhancement of the critical currents as well potential applications in future fluxtronic 
devices (see e.g. Refs. \cite{Bulaevskii,Gardner,Goa,Helseth1,Helseth2,Kayali,Daumens} and references therein).
So far the theoretical approaches developed have assumed hard nanomagnets, i.e. magnets that are 
not influenced by the vortex field (see e.g Refs. \cite{Helseth2,Kayali} and references therein). However, in many cases the magnetic film is 
soft\cite{Goa}, and the vortices will therefore induce a magnetization distribution in these films. The purpose of 
this paper is to make a first simple approach to study the vortex interactions in presence of a soft magnetic film
that is homogenous in absence of vortices.

To this end, we consider a simple model based on a thin superconducting film of infinite extent, located at 
z=0 with thickness $d$ much smaller than the penetration depth of the
superconductor. The surface is covered by a soft magnetic film of thickness smaller than 
that of the superconducting film. We also assume that there is no spin diffusion and 
proximity effects. In general, the current density is a sum of the supercurrents and magnetically 
induced currents, which can be expressed through the generalized London 
equation as\cite{Helseth2} 
\begin{equation}
\mbox{\boldmath $\nabla \times J$} = -\frac{1}{\lambda ^{2}} \mbox{\boldmath $H$} + 
\frac{1}{\lambda ^{2} } V(\mbox{\boldmath $\rho$}) \delta (z) \mbox{\boldmath $\hat{e}$}_{z} + 
\mbox{\boldmath $\nabla \times \nabla \times M$}\,\,\, ,
\label{LA}
\end{equation}
where $\mbox{\boldmath $J$}$ is the current density, $\lambda$ is the penetration depth, $\mbox{\boldmath $H$}$ is the 
magnetic field and $\mbox{\boldmath $\nabla$} \times \mbox{\boldmath $M$}_{V}$ 
is the magnetically induced current. Note that the magnetically induced currents are included as the last term on the 
right-hand side of Eq. \ref{LA}, and are therefore generated in the same plane as the vortex. This is justified since 
we assume that the thicknesses of both the superconducting and magnetic films are smaller than the penetration depth, and 
the magnetic film is located very close to the superconductor.
The vortex is aligned in the z direction, and its source function $V(\mbox{\boldmath $\rho$})$ is assumed to
be rotational symmetric. In the case of a Pearl vortex we may set 
$V(\rho) =(\Phi _{0} /\mu _{0}  )\delta (\rho)$, where $\Phi _{0}$ is the flux quantum and 
$\mu _{0}$ the permeability of vacuum. 

Let us assume an initially homogenous soft magnetic film without domain walls, consisting of a single domain 
with in-plane magnetization in absence of external magnetic fields. Moreover, we assume that the free energy 
of this domain can be expressed as a sum of the uniaxial anisotropy 
and the demagnetizing energy. The magnetic field from a vortex tilts the magnetization vector out of the plane 
due to the additional Zeeman energy, and it can be shown that for small tilt angles one has \cite{Helseth1} 
\begin{equation}
\mbox{\boldmath $M$} =\left( M_{\rho},M_{z} \right) \approx \left(M_s,M_s\frac{H_{z}}{H_{a}} \right) \,\,\, ,
\label{tilt}
\end{equation}
where $M_{\rho}=\sqrt{M_{x}^{2}+M_{y}^{2}}$ and $M_{z}$ are the in-plane (radial) and perpendicular components of the magnetization, 
$H_{\rho}=\sqrt{H_{x}^{2} +H_{y}^{2}}$ and $H_{vz}$ are the in-plane (radial) and perpendicular components of the vortex field, and 
$M_{s}$ is the saturation magnetization of the magnetic film. $H_{a}=M_{s} -2K_{u}/\mu_{0}M_{s}$ is the socalled anisotropy field, where 
$K_{u}$ is the anisotropy constant of the magnetic film. Note that since we neglect the cubic anisotropy of the system, 
the in-plane magnetization direction must be the same as that induced by the vortex field. It is important to point out 
that the total magnetization vector $\mbox{\boldmath $M$} $ is not directed along the vortex field, since there is always a 
large component of the magnetization in the plane of the magnetic film due to the uniaxial anisotropy.    
Upon using Eq. \ref{tilt}, we neglect the contribution from the exchange energy, which is justifiable for sufficiently 
large $K_{u}$ or small spatial field gradients.

In order to solve the generalized London equation, we follow the method of Ref. \cite{Helseth2}.
The current can only flow in a thin layer of thickness 
$d$ ($d\ll\lambda$), and it is therefore necessary to average over this thickness and 
consider the z component only
\begin{equation}
H_{z} + \lambda _{e} \left( \frac{\partial J^{s}_{y}}{\partial x}
-\frac{\partial J^{s}_{x}}{\partial y} \right) =V(\rho ) +d\lambda _{e}\left( \mbox{\boldmath $\nabla \times \nabla \times M$} \right) _{z} \,\,\, ,
\label{A}
\end{equation}
where $\lambda _{e} =\lambda ^{2} /d$ is the effective penetration depth. 
The sheet current flowing in the thin layer is now given by $\mbox{\boldmath $J$}_{s}=d\mbox{\boldmath $J$}$.
The Maxwell equation $\mbox{\boldmath $\nabla \times H$}=\mbox{\boldmath $J$}$
gives
\begin{equation}
J_{x} = \frac{\partial H_{z}}{\partial y}
-\frac{\partial H_{y}}{\partial z},   \,\,\,\,\,\,\,\,\,
J_{y} = \frac{\partial H_{x}}{\partial z}
-\frac{\partial H_{z}}{\partial x}\,\,\, .
\end{equation}
Since all derivatives $\partial /\partial z$ are large compared to 
the tangential $\partial /\partial \mbox{\boldmath $\rho$}$, we may set
\begin{equation}
J^{s}_{x} \approx H^{-}_{y} - H^{+}_{y},   \,\,\,\,\,\,\,\,\, 
J^{s}_{y} \approx H^{+}_{x} - H^{-}_{x} \,\,\, ,  
\end{equation}
where $H^{+}_{i}$ and $H^{-}_{i}$ (i=x, y) are the components at the upper and
lower surfaces, respectively. Since the environments of the upper and lower
half-spaces are identical, we have the boundary condition 
$H^{+}_{i} =-H^{-}_{i}$, which results in
\begin{equation}
J^{s}_{x} \approx  -2H^{+}_{y},  \,\,\,\,\,\,\,\,\,  J^{s}_{y} \approx 
2H^{+}_{x} \,\,\, .
\end{equation} 
Using $\mbox{\boldmath $\nabla \cdot H$} =0$, Eq. (\ref{A}) becomes
\begin{equation}
H_{z} - 2\lambda _{e} \frac{\partial H_{z}}{\partial z} = 
V (\rho) +d\lambda _{e}\left( \mbox{\boldmath $\nabla \times \nabla \times M$} \right) _{z} \,\,\, .
\label{Lond}
\end{equation}

Using Eq. \ref{tilt} and the fact that our system has no volume charges, we find  
\begin{equation}
H_{z} - 2\lambda _{e} \frac{\partial H_{z}}{\partial z} + \alpha \left(\frac{\partial ^{2}H_{z}}{\partial x^{2}} + \frac{\partial ^{2}H_{z}}{\partial y^{2}} \right) = 
V(\rho ) \,\,\, ,  \,\,\,  \alpha = \frac{d\lambda _{e}M_{s} }{H_{a} }      \,\,\, .
\label{vortexsolution}
\end{equation}
 
In order to solve Eq. (\ref{vortexsolution}), it is useful to note that our system is rotational symmetric, 
and also that $\mbox{\boldmath $\nabla \times H$}=\mbox{\boldmath $\nabla \cdot H$}=0$ 
outside the system. Therefore, we can introduce a scalar potential $\phi$ which
vanishes at $z\rightarrow \pm \infty$
\begin{equation}
\phi (\mbox{\boldmath $\rho$}, z) = \frac{1}{(2\pi)^{2}}
\int_{-\infty}^{\infty} \int_{-\infty}^{\infty}  \phi( \mbox{\boldmath
$k$} ) \exp(i \mbox{\boldmath $k \cdot \rho$} -k|z|) d^{2}k\,\,\, ,
\label{F}
\end{equation}
where $k=\sqrt{k_{x}^{2} +k_{y}^{2}}$, and $\mbox{\boldmath
$H$}=\mbox{\boldmath $\nabla$} \phi$. It is helpful to note that
$H_{z}(\mbox{\boldmath $k$})=-k\phi (\mbox{\boldmath $k$})$ for the upper
half-space. 

Applying the Fourier transform to Eq. (\ref{vortexsolution}), we obtain 
\begin{equation}
\phi (k) =-\frac{V(k)}{(1+2\lambda _{e}k -\alpha k^{2})k}
\,\,\, .
\end{equation}
We assume that the vortex is located at the origin, and therefore the resulting scalar potential is
\begin{equation}
\phi (\rho,z) =-\frac{1}{(2\pi)^{2}}\int_{-\infty}^{\infty} 
\int_{-\infty}^{\infty} \frac{V(k) \exp(i \mbox{\boldmath $k \cdot \rho$}
-k|z|)}{k(1+2\lambda _{e}k -\alpha k^{2})} d^{2}k
\,\,\, .
\end{equation}
Due to the rotational symmetry the potential is found to be
\begin{equation}
\phi (\rho ,z) =-\frac{1}{2\pi }\int_{0}^{\infty} 
V(k) \frac{J_{0} (k\rho)}{1+2\lambda _{e}k -\alpha k^{2}} \exp(-k|z|)dk
\,\,\, ,
\end{equation}
where we have used that
\begin{equation}
\int_{0}^{2\pi} \exp(ik\rho cos\phi )d\phi =2\pi J_{0} (k\rho) \,\,\, .
\end{equation}
To obtain the magnetic field components (in the radial and z direction), we apply the following formula
\begin{equation}
\frac{d}{d\rho} J_{0} (k\rho )=-kJ_{1} (k\rho ) \,\,\, , 
\end{equation}
and find 
\begin{equation}
H_{z} (\rho,z) =\frac{1}{2\pi }\int_{0}^{\infty} 
kV(k)\frac{J_{0} (k\rho)}{1+2\lambda _{e}k -\alpha k^{2}} \exp(-k|z|)dk
\,\,\, ,
\end{equation}
\begin{equation}
H_{\rho} (\rho,z) =\frac{1}{2\pi }\int_{0}^{\infty} 
kV(k)\frac{J_{1} (k\rho)}{1+2\lambda _{e}k -\alpha k^{2}} \exp(-k|z|)dk
\,\,\, .
\end{equation}
In the case $\alpha =0$ and $V(k) =\Phi_{0}/\mu_{0}$, the 
field is reduced to that of the standard Pearl solution\cite{Pearl,Helseth2}.
We also notice that there is a divergency in k-space when $1+2\lambda _{e}k -\alpha k^{2} =0$. This is due to 
the fact that we assumed a very thin superconductor ($d\ll\lambda$). The divergency will therefore smoothen out upon 
solving the London equation for arbitrary thicknesses $d$, but this task is outside the scope of the current work. 
Nonetheless, it is seen that at small distances (large $k$) the magnetic field may change sign as 
compared to the standard Pearl solution.
On the other hand, at large distances (small $k$) the field is basically not influenced by the soft magnetic 
film. A typical scale for the crossover is $k\sim H_{a}/M_{s}d \sim 1/d$, since in many practical cases 
$M_{s}\sim H_{a}$. We argue that the crossover follows this dependence also if we allow for a thicker 
magnetic film, but at some point one must take into account the magnetic volume charges. 

Let us now try to estimate the interaction energy and force between two vortices. To this end, we note that 
there are two contributions to this interaction. First, the energy associated with the two 
vortices (with indices $1$ and $2$, respectively) is given by
\begin{equation}
E_{v} = \mu _{0}d\int _{-\infty}^{\infty} \int _{-\infty}^{\infty} \mbox{\boldmath $H$}_{1} \mbox{\boldmath $\cdot$} \mbox{\boldmath $V$}_{2}d^{2}\rho \,\,\, ,
\end{equation}
which for Pearl vortices is found to be
\begin{equation}
E_{v}(\rho) = \Phi_{0} H_{z} (\rho)d =\frac{\Phi_{0} ^{2}d}{2\pi \mu_{0}}\int_{0}^{\infty} 
k\frac{J_{0} (k\rho)}{1+2\lambda _{e}k -\alpha k^{2}} dk \,\,\, .
\end{equation}
One should also take into account the interaction between the vortex field and the vortex-induced 
magnetization, which can be found by using
\begin{equation}
E_{m} = - d\int _{-\infty}^{\infty} \int _{-\infty}^{\infty} \mbox{\boldmath $M$}_{1} \mbox{\boldmath $\cdot B$}_{2} d^{2}\rho  \,\,\, .
\label{magneti}
\end{equation}
Note that the magnetic induction in the magnetic film is $\mbox{\boldmath $B$}_{2} = \mu _{0}(\mbox{\boldmath $H$}_{2} +\mbox{\boldmath $M$}_{2})$, 
which means that to the first order $E_{m}$ is a constant proportional to $\mu _{0}M_{s}^{2}$. It should be emphasized 
that this is only correct when the vortex field is substantially weaker than the anisotropy field and magnetization of the magnetic film. 

Based on the above observations we estimate the force between two vortices to be 
\begin{equation}
F(\rho) \approx \frac{\Phi_{0} ^{2}d}{2\pi \mu_{0}}\int_{0}^{\infty} 
k^{2}\frac{J_{1} (k\rho)}{1+2\lambda _{e}k -\alpha k^{2}} dk \,\,\, .
\end{equation}
When the distance is large ($\rho \gg \lambda _{e}$), the soft magnetic film does not influence the vortex 
interaction, and the force is therefore repulsive and governed by the standard Pearl solution\cite{Pearl}. 
On the other hand, at smaller distances (still larger than the coherence length $\xi$), we may 
approximate the force by
\begin{equation}
F(\rho) \approx -\frac{\Phi_{0} ^{2}d}{2\pi \mu_{0} \alpha  \rho}\,\,\, , \,\,\,  \xi < \rho \ll \frac{M_{s}}{H_{a}} d \,\,\, , 
\label{f}
\end{equation}
where we have used that $\int_{0}^{\infty} J_{1} (k\rho) dk =1/\rho$.
Therefore, at small distances $\rho$ the force is attractive in presence of a soft magnetic film, which is quite 
surprising. However, one can interpret this as a result of the currents generated by the vortex-induced 
magnetization. These currents generate a magnetic field which opposes that produced by the vortex in absence of 
a magnetic film, and this field interacts with the core of the second vortex.
For comparison, we note that in absence of a soft magnetic film the force at small distances is given by
\begin{equation}
F(\rho) \approx \frac{\Phi_{0} ^{2}d}{4\pi \mu_{0} \lambda _{e} \rho ^{2}}\,\,\, , \,\,\,  \xi < \rho \ll \lambda _{e} \,\,\, , 
\end{equation}
which is seen to be repulsive with a magnitude that is decaying faster with increasing $\rho$ than Eq. \ref{f}.

One may expect that the presence of a soft magnetic film results in new vortex 
configurations and also influences the creation of vortices at the superconducting 
transition temperature. We therefore hope that this study will stimulate further work in 
this field.

\end{document}